# Large longitudinal and anomalous transverse Magneto-thermoelectric effect in kagome antiferromagnet FeGe

Jiajun Ma[1,#], Rong Chen[5,8,#], Yazhou Li[1], Chenfei Shi[2], Yantao Cao[3], YuWei Zhang[1], Jiaxing Liao[1], Yunfei Han[1], Guangxi Wen[1], Jialu Wang[1], Hanjie Guo[3], Jianhui Dai[1], Chenguang Fu[4,7], Jin-Ke Bao[1,$], Yan Sun[5, &], Zhu-An Xu[6,7, #], Yuke Li[1,*]

1 *School of Physics and Hangzhou Key Laboratory of Quantum Matters, Hangzhou Normal University, Hangzhou 311121, China*
2 *Department of Physics, Materials Genome Institute and International Center for Quantum and Molecular Structures, Shanghai University, Shanghai 200444, China*
3 *Songshan Lake Materials Laboratory, Dongguan, Guangdong 523808, China*
4 *School of Materials Science and Engineering, Zhejiang University, 310058 Hangzhou, China*
5 *Shenyang National Laboratory for Materials Science, Institute of Metal Research, Chinese Academy of Sciences, Shenyang, 110016, China*
6 *School of Physics, Zhejiang University, Hangzhou 310027, China*
7 *State Key Laboratory of Silicon and Advanced Semiconductor Materials, Zhejiang University, 310058 Hangzhou, China*
8 *School of Materials Science and Engineering, Northeastern University, Shenyang 110819, China*

## ABSTRACT

**Topological Kagome magnets, characterized by nontrivial electronic band structures featuring flat band, Dirac cone and van Hove singularities, provide a new avenue for the realization of thermoelectric devices. Unlike the conventional longitudinal Seebeck effect, transverse thermoelectric (TE) effects like the Nernst effect have attracted growing interest due to their unique transverse geometry and potential advantages. Here, we report the observation of a significant transverse thermoelectric conductivity $\alpha^A_{zx}$ of 15 A K$^{-1}$m$^{-1}$ at low temperatures, together with a pronounced anomalous Nernst effect in the Kagome antiferromagnet FeGe, which exhibits a charge density wave inside the antiferromagnetic (AFM) state. This value is the highest record among known AFM materials. Furthermore, the thermopower at 14 T increases by $10^2$-$10^4$% around the canted-AFM (CAFM) transition temperature, $T_{cant}$, comparable to that of the well-known AFM thermoelectric materials. These effects are attributed to large Berry curvature arising from the non-collinear spin texture in FeGe, highlighting its potential for enhancing thermoelectric performance and its candidacy for magneto-TE applications in Kagome antiferromagnetic materials.**

## 1. Introduction

Over the past few decades, many studies on traditional thermoelectric (TE) materials have mainly focused on the longitudinal Seebeck effect to improve TE performance by band tuning and phonon engineering, such as carrier density, defects, and impurities.[1-5]. Significant progress has been made in the study of superior TE materials and their enhanced performance[6]. By contrast, the investigation of novel electronic structures like the Dirac cone and Berry curvature opens up new avenues for advancing thermoelectric research. The so-called topological materials discovered recently hold promise for realizing large TE performance. In particular, the transverse magneto-TE effect holds significant potential for applications in waste-heat harvesting and solid-state cooling[7-13]. Furthermore, the transverse thermoelectric benefits of anomalous Nernst effect (ANE), including simplified junction design and the ability to develop thin-film devices[10-12], offer advantages such as scalability, suitability for mass production, and enhanced flexibility[13].

Recently, extensive research efforts on the anomalous Nernst effect (ANE) in magnetic topological materials(MTMs)[7, 14-31] have attracted increasing attention because of its potential advantages. First, the transverse Nernst devices, which relate the perpendicular electric fields to temperature gradients, can be easily fabricated in a single material without complicated electrical connections and contact resistances, compared with traditional thermoelectric devices. Second, the Nernst effect is not limited to single- or multi-band carrier transports[9]. A multiband system with hole and electron pockets can substantially enhance transverse TE performance[9, 13]. Take two types of typical topological magnetic materials as examples: One is the topological ferromagnets such as $Co_2MnGa$[21, 25, 30, 32-34], $Co_3Sn_2S_2$[18, 19], $Fe_3Sn$[17], and $Fe_3Sn_2$[15, 24]. They exhibit large anomalous Nernst signals, reaching values as high as 1~ 8 $\mu V/K$, which significantly exceed those of conventional ferromagnets that typically correlate with their magnetization. The intrinsic ANE generally originates from the large Berry curvature in the momentum space due to topological band crossings (Weyl nodes) and anti-crossings in the electronic structures[18, 19, 22, 23]. Another example is the antiferromagnets with peculiar noncollinear/noncoplanar spin structures, for example, $Mn_3Sn/Ge$[20], $YbMnBi_2$[35], and $YMn_6Sn_6$[14]. They have a zero or very small net magnetization, but the ANE can be observed owing to the nontrivial spin textures. Therefore, topological materials are believed to be excellent platforms for exploring transverse as well as longitudinal TE. However, the study of high-performance magneto-TE materials is still scarce. Furthermore, how spatially varying spin textures may influence TE transport in strongly correlated electronic systems is still less reported. Exploring the interplay of charge, topology, and magnetism to improve the TE performance in MTMs provides new perspectives for the application of TE technology [see Figure 1(a)].

Very recently, a topological Kagome material FeGe has triggered significant interest because it is a strongly correlated kagome metal and has a short-range charge density wave (CDW) phase ($T_{CDW}$ ~ 100 K) deeply inside the A-type AFM phase occurring at $T_N$ ~ 400 K[36]. At a lower temperature of $T_{cant}$ ~ 60 K, its magnetic moments form a canted double-cone AFM structure with an interlayer turn angle for the basal-plane moment component[37, 38]. The neutron scattering experiment found a significant enhancement of Fe moments in the presence of CDW order[36, 39], indicating the intertwined magnetic and charge orders in the kagome metal FeGe[40-42]. Both scanning tunneling microscopy (STM)[43, 44] and angle-resolved photoemission spectroscopy (ARPES)[41] supported the short-range 2 × 2 × 2 CDW ordering in as-grown FeGe samples, whereas, other experiments suggested the formation of long-range CDW ordering in the annealed samples [40, 45-47], and its coexistence with a pronounced edge state[44], indicating a strong entanglement of topology, charge order, and magnetism in the kagome FeGe. However, such unique features in FeGe make it an excellent platform for the realization of anomalous transport properties[31], in particular the large magneto-TE effect. In addition, magnetic-field-induced spin-flop transition from the canted AFM to the non-collinear AFM order can give birth to a nontrivial Berry phase in real space, which may imply promising high transverse TE performance.

Here we report the observation of a magnetic-field-induced large change in thermopower, an anomalous Nernst effect, and one of the largest transverse TE conductivities in FeGe. At the critical temperature $T_{cant}$ associated with the entrance of the canted AFM order, several transport coefficients including the Hall conductivity $\sigma_{zx}$, thermopower $S_{zz}$ and Nernst signal $S_{zx}$ change their sign from positive to negative. The relative change in thermopower reaches approximately **$10^2$-$10^4$%** around $T_{cant}$ and 14 T, the largest value for the magneto-power in the well-known topological AFM. More importantly, we observed a large ANE below $T_{cant}$, with a maximum $S^A_{zx}$ of 1.2 $\mu$V/K at 20 K which is higher than that of the Kagome canted AFM $Mn_3Sn/Ge$. As a result, a large anomalous transverse TE conductivity $\alpha^A_{zx}$ of 15 AK$^{-1}$m$^{-1}$ is obtained. This is the largest value observed for AFM materials to date, which is also comparable to the report for a ferromagnet $UCo_{0.8}Ru_{0.2}Al$[23]. Thus, our results demonstrated that the nontrivial spin structures coupling with charge order as well as topological band in a material may produce a large contribution of Berry curvature to the large $\alpha^A_{zx}$.

## 2. Results and Discussion

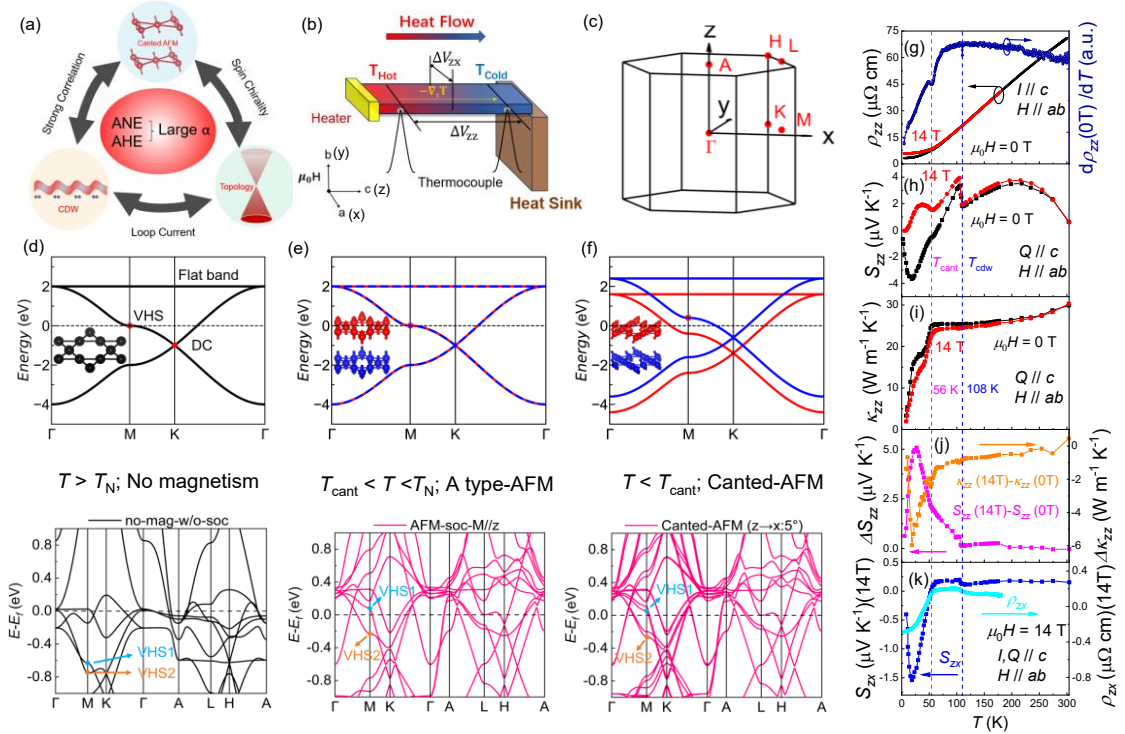

Figure 1 (a) A cartoon illustration of the interaction among canted AFM, CDW and topology in a kagome AFM, which may induce ANE and AHE in FeGe owing to the nontrivial Berry phase, yielding a large TE conductivity. (b) Schematic picture of TE measurements including Seebeck and Nernst effects. (c) The Brillouin zone (BZ) and high-symmetry points of P6/mmm space group (No. 191). (d-f) The evolution of electronic band structure in Kagome lattice with including inter-layer antiferromagnetic order and in-plane spin canting. The top and bottom panels represent the schematic and realistic materials of FeGe, respectively. (g) The longitudinal resistivity $\rho_{zz}$ as a function of temperature at zero and 14 T as I// c and B//ab. (h) Temperature dependence of thermopower and thermal-conductivity (i) at zero and 14 T (B// ab-plane) in FeGe sample. (j) The magneto-thermopower, $\Delta S_{zz}$ = ($S_{zz}$(14T) − $S_{zz}$(0)), and thermal-conductivity $\Delta k_{zz}$ = ($k_{zz}$(14T) − $k_{zz}$(0)) vs. temperature. (k) The Hall resistivity $\rho_{zx}$ and Nernst signal as functions of temperature at 14 T.

Using the configuration shown in Fig. 1(b), we performed the longitudinal resistivity, thermopower, thermal-conductivity, Hall resistivity, and the Nernst coefficient measurements. Temperature dependence of resistivity $\rho_{zz}$ (or $\rho_c$, where the current flows in the c-direction), as shown in Figure 1g, indicates two consecutive phase transitions, including a CDW transition at 108 K and a canted-AFM magnetic transition at 56 K, similar to those in the previous report[36]. The applied magnetic fields do not change the $T_{cdw}$, but slightly shift the $T_{cant}$ to lower temperature, implying the robust charge order. Figure 1d shows the typical electronic band structure of a dimensional (2D) Kagome lattice, featuring one flat band in the whole Brillouin zone (BZ), one Dirac cone at K point, and two van Hove singularities near M point. In FeGe, two layers of Kagome lattice formed by atom Fe can form inter-layer antiferromagnet at low temperature, with spin degeneration at each k point, see Figure 1e. Near the transition temperature, in-plane spin canting appears, the spin degeneracy is broken, and each spin channel keeps the main features, as presented in Figure 1f. Owing to the typical topological band structure in Kagome lattice and spin canting, strong thermoelectrical responses are expected.

The thermopower proportionally depends on the logarithmic derivative of the Density of States (DOS) for energy at the Fermi level and is regarded as a sensitive probe for the changes in the Fermi surfaces, such as CDW ordering and structural distortions. Figure 1h shows the thermopower $S_{zz}$ (both temperature gradient and current flow in the c-directions) as a function of temperature at zero and 14 T. It provides fruitful physical information about the interplay between charge order and magnetism in FeGe. Under zero-field conditions, the thermopower is positive at high temperatures, then starts to increase with decreasing temperature, and exhibits a large broad hump around 200 K with its value of 3.8 $\mu$V/K. Following that, a prominent jump at $T_{cdw}$ associated with the CDW transition can be observed. The enhanced thermopower below $T_{cdw}$ can be ascribed to a decrease in the density of states due to the partial opening of the gap around the Fermi surface. Upon further decreasing temperature to $T_{cant}$, a detectable anomaly can be found, which may be caused by the increase in magnetic scatterings due to the transition from collinear AFM to the canted AFM order[36]. Coincidentally, the $S_{zz}$ crosses zero from positive to negative at $T_{cant}$, and continues to decrease down to approximately 4 $\mu$V/K at 25 K and oppositely increases with temperature further cooling down to 5 K. The sign-changed $S_{zz}$ indicates a change of majority charge carrier from hole to electron around $T_{cant}$.

Applying a magnetic field, the substantial differences in thermopower can be observed in Figure 1h. The applied magnetic field of 14 T induces a completely positive thermopower in the whole temperature regime. Based on the Mott relation $S_{zz} = \frac{\pi^2}{3} \frac{k_B^2 T}{e} \frac{\partial \sigma_{zz}}{\partial \mu}$, where $\sigma_{zz}$ is the longitudinal conductivity and $\mu$ denotes the chemical potential, this may be ascribed to a field-induced shift of the chemical potential $\mu$ [48]. Although the magnetic field does not change $T_{cdw}$ too much, it can obviously increase the thermopower in the CDW state and significantly enhance the magnitude of the anomaly below $T_{cant}$. Meanwhile, the valley of the negative $S_{zz}(T)$ at zero field becomes a positive peak in $S_{zz}(T)$ under 14 T. Thus, the difference $\Delta S_{zz} = S_{zz}(14T) - S_{zz}(0T)$ associated with the contribution of magneto-power starts to increase substantially below $T_{cdw}$ and peaks around 25 K in Figure 1j, implying the strong correlation between CDW and magnetic order in FeGe. Such features in thermopower are rarely observed in a strong correlation system. The large magneto-thermopower effects in materials are attributed to the shift of chemical potential in the linear bands by magnetic field. Similar behaviors can be found in the thermal-conductivity $k_{zz}$, as shown in Figure 1i. The $k_{zz}$ at zero field shows a detectable increase at $T_{cdw}$ and a sharp drop at $T_{cant}$, as well as a broad hump around 15 K. Applying a magnetic field of 14T can suppress $k_{zz}$ below $T_{cdw}$. As a result, a large divergence between $k_{zz}(14T)$ and $k_{zz}(0T)$ is observed, similar to the case in thermopower. The obtained difference $\Delta k_{zz} = k_{zz}(14T) - k_{zz}(0T)$ is plotted in Figure 1i. The large peak in $\Delta k_{zz}$ around 25 K supports the contribution of magneto-thermal-conductivity associated with the canted AFM order, consistent with the $\Delta S_{zz}$. Note the characteristic peak around 25 K is also found in the Nernst coefficient $S_{zx}$, but is absent in the Hall resistivity $\rho_{zx}$ in Figure 1k. This means that only thermal-transport coefficients can reflect the characteristic peak, indicating a change of magneto-entropy related to the electrons near the Fermi level.

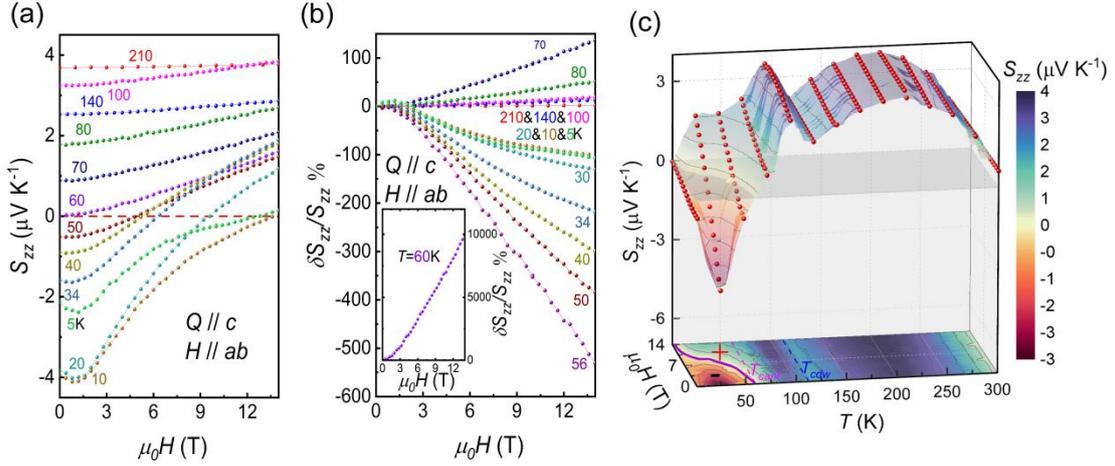

Figure 2 (a) Thermopower as a function of the magnetic field at different temperatures, where $S_{zz}$ is negative below $T_{cant}$ and becomes positive above $T_{cant}$ at zero fields. (b) The field dependence of the change of thermopower. (c) The phase diagram of thermopower vs. both magnetic fields and temperature.

We further investigated the magnetic field dependence of thermopower $S_{zz}$ and magneto-thermopower ratio $\Delta S_{zz}/S_{zz} = |(S_{zz}(H) - S_{zz}(0))/S_{zz}(0)|$ for FeGe at different temperatures, as shown in Fig. 2a-2b. Above $T_{cdw}$, $S_{zz}$ does not change too much under the magnetic fields. At $T_{cant} < T < T_{cdw}$, the positive $S_{zz}$ increases linearly with the increase of the magnetic field up to 14 T similar to the linear field dependent magnetoresistance. The $\Delta S_{zz}$ gradually increases with decreasing temperature and reaches the maximum $\Delta S_{zz}/S_{zz} \sim 10^4$ % at 14 T and 60 K (see detailed information in SI). Such large magneto-thermopower is less reported. This value is one or two orders of magnitude larger than those reports in manganite [49] (~ 1400%), conventional thermopower materials $Ag_{2-\delta}Te$ (~

500%)[50] and other topological semimetals MnSi(Ge)[51], YMn$_6$Sn$_6$[14], Yb/EuMnBi(Sb)$_2$[35, 52], although the value of $S_{zz}$ is rather small (~ 4 μV/K). At $T < T_{cant}$, the $S_{zz}$ becomes negative at zero-field, tends to decrease and exhibits a minimum valley at low fields, and then begins to increase toward positive values and bend at high fields with temperature cooling down, accompanied by a decrease in $\Delta S_{zz}$ of ~ 150 % at 5 K. The characteristics of thermopower below $T_{cant}$ can be partially attributed to the significant entropy change linked to the crossover of charge dynamics from a coherent to an incoherent state during magnetic structural changes. Both thermal conductivity and the Nernst effect (discussed in Figure 3) show the same behaviors at the characteristic temperature. Note that the slope-change of $\Delta S_{zz}/S_{zz}$ at $T_{cant}$ also indicates the strong coupling between thermoelectric properties and magnetic structures. The thermopower as functions of magnetic fields and temperature is plotted in Figure 2c. Below $T_{cant}$, the sign of thermopower and its magnitude depend strongly on the intensity of magnetic fields, suggesting the intimate coupling of spin and charge order in FeGe[42].

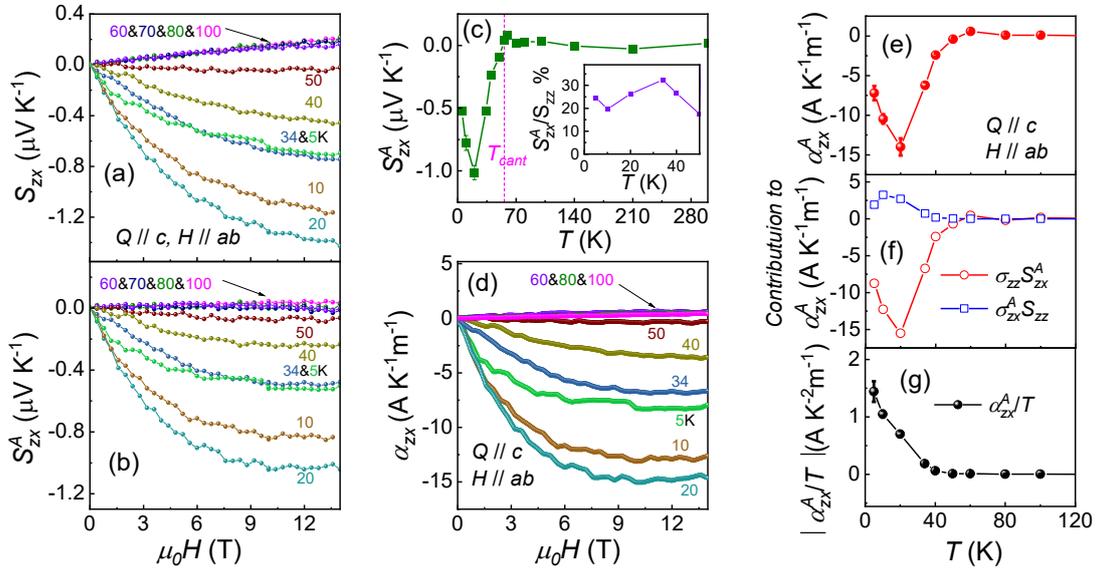

Figure 3 The anomalous Nernst effect. (a) The Nernst signal $S_{zx}$ as a function of magnetic fields up to 14 T at different temperatures. (b) The anomalous Nernst signal $S^A_{zx}$ vs. magnetic fields. (c) The extracted $S^A_{zx}$ as a function of temperature. The inset shows the Nernst angle below $T_{cant}$. (d) The calculated transverse thermoelectric conductivity, $\alpha_{zx}$, as a function of magnetic fields using the measured, $\sigma_{zz}$, $\sigma_{zx}$, $S_{zz}$, and $S_{zx}$ data. Temperature dependence of the extracted $\alpha^A_{zx}$ (e), two components $\sigma_{zz}S^A_{zx}$ and $\sigma^A_{zx}S_{zz}$ (f), and $\alpha^A_{zx}/T$ (g).

In addition, the compounds with canted AFM spin structures[20, 35] have been found to exhibit exotic physical phenomena, such as the anomalous Hall effect and Nernst effect. The magnetic field dependence of the Nernst signal is shown in Figure 3a-3b. Similar to the thermopower, the transverse Nernst signal $S_{zx}$ is positive and exhibits a linear field-dependence above $T_{cant}$, and it would like to approach zero and remain less changed under magnetic fields around $T_{cant}$. As $T < T_{cant}$, the $S_{zx}$ as a function of magnetic field starts to bend and tends to saturate at high fields, resembling the character the of Nernst effect in the antiferromagnet YbMnBi$_2$[35]. The maximum $S_{zx}$ reaches approximately 1.5 μV/K, which is also comparable to values observed in most of topological magnets, such as Mn$_3$Sn[20], Fe$_3$Sn$_2$[15, 24], Co$_3$Sn$_2$S$_2$ [18, 19], RMn$_6$Sn$_6$-family[14, 16]. To extract the anomalous contribution ($S^A_{zx}$), one can separate it from the total Nernst signal by subtracting a local linear ordinary Nernst effect at high fields, as shown in Figure 3b. The anomalous Nernst signal ($S^A_{zx}$) was thus obtained by extrapolating the slope to a zero-field value, and the extracted $S^A_{zx}$ as a function of temperature is plotted in Figure 3c. The finite $S^A_{zx}$ appears below $T_{cant}$, and peaks around 25 K. The large $S^A_{zx}$ thus denotes a large Nernst angle with the maximum value of 30% displayed in the inset of Figure 3c.

The transverse thermoelectric conductivity ($\alpha_{zx}$) (Figure 3d) can be calculated using our measured $\sigma_{zz}$, $\sigma_{zx}$, $S_{zz}$, and $S_{zx}$ according to Equation: $\alpha_{zx} = \sigma_{zz}S_{zx} + \sigma_{zx}S_{zz}$. $\alpha_{zx}$ shows a similar field-dependent behavior as that of the Nernst signal $S_{zx}$. It increases linearly with increasing field above $T_{cant}$, but as $T < T_{cant}$, $\alpha_{zx}$ presents a large steplike feature with a saturated value at high fields, implying the anomalous TE effect. We thus extracted the anomalous $\alpha^A_{zx}$ from the $\alpha_{zx}$ by extrapolating the high fields data to the zero-field value in Figure 3e. As a comparison, we also calculated the $\alpha^A_{zx}$ using the sum of two contributions $\sigma_{zz}S^A_{zx}$ and $\sigma^A_{zx}S_{zz}$ in Figure 3f. The former component is almost one order of magnitude larger than the latter one at 20 K, dominating the $\alpha^A_{zx}$. This suggests that the large electrical conductivity and relatively small thermopower play a crucial role in enhancing the $\alpha_{zx}$. This outstanding feature of FeGe is distinct from most known topological materials which are determined by $\sigma^A_{zx}S_{zz}$[17-19]. The $|\alpha^A_{zx}|$ peaks at 20 K with a value of ~15 AK$^{-1}$m$^{-1}$, which is a relatively large value, especially for an antiferromagnet. It is also much higher than most of the topological ferromagnets, including Co$_3$Sn$_2$S$_2$[18, 19], Co$_2$MnGa[30], Fe$_3$Sn[17], and Fe$_3$Sn$_2$ [15] with the record value of 1~6 AK$^{-1}$m$^{-1}$ and one or two orders of magnitude higher than those of the antiferromagnets Mn$_3$X (X = Sn and Ge)[20] and YMn$_6$Sn$_6$[14]. Last, considering the Mott relation, the unconventional large $\alpha^A_{zx}/T$ appears below $T_{cant}$ in Figure 3g indicates that the non-trivial spin texture in the entrance of CAFM state may appropriately modify the band structure near Fermi level, resulting in a large curvature in the moment space.

To investigate the relation between canted spin structure and ANE, we performed first-principles calculations in CAFM states. Owing to a complex of realistic magnetic structures, we only chose some specific canted angles and different interlayer relations, with a maximum unit cell containing 12 Fe atoms, see Figure S13 for the details. The calculated anomalous Nernst conductivity (ANC) value in the given CAFM states lies in a large range from 0 to ~1 AK$^{-1}$m$^{-1}$, dependent on the details of magnetic structures. The ANC can reach up to around 3.5 AK$^{-1}$m$^{-1}$ with a weak electron doping to move up the Fermi level to around 0.3 eV. We also calculated the case with fully in-plane spin polarization under the situation of strong in-plane magnetic field. Similar to the CAFM states, the ANC in collinear FM state is around 0.2 A K$^{-1}$ m$^{-1}$ and 2.0 AK$^{-1}$m$^{-1}$ near neutrality point and E$_0$+0.3 eV, respectively, as presented in Figure S14. Since we cannot capture all the possible magnetic structures in CAFM states, a much larger ANC might exist.

The appearance of ANE only in the CAFM state implies that the nontrivial spin structures may play an important role, although the CDW order remains largely unaltered. Note the saturated $S_{zx}$ at low temperature is only observed as $B$ is over 6 T (in Figure 3b), which accidentally meets the critical field of spin-flop transition along the c-direction. During the process of SF transition, the CAFM state into a noncollinear/coplanar spin arrangement can give birth to a large nontrivial Berry phase in moment space.

To uncover the origin of ANE, we examine the experimental data above. Firstly, we found that both the $S_{zz}$ and the $S_{zx}$ undergo the sign-change from positive to negative at $T_{cant}$, manifesting the dominant electron carriers and the shift of chemical potential in the CAFM state. In a metal, the heat transport phenomena only involve the electrons/holes within their energy range approximately between $\varepsilon_F \pm k_B T$ due to the Fermi distribution function. If the band structures around fermi energy $\varepsilon_F$ are symmetric, the entropy as well as the potential energy with opposite signs produced by the heat flow of electrons/holes below and above $\varepsilon_F$ will cancel out each other, resulting in a negligible thermopower. A tiny difference in the number, velocity, and scattering rate of carriers can break such cancellation. In FeGe, the Density Functional Theory (DFT) calculation[41] has demonstrated the sharp change of band structure around $\varepsilon_F$ in the CAFM state. Therefore, only two types of carrier models do not explain the sign-change in thermopower. In addition, the transverse Nernst signal $S_{zx}$ usually tends to vanish in a single-band metal because of the Sondheimer cancellation[53]. These unusual features where both $S_{xx}$ and $S_{zx}$ simultaneously cross zero near $T_{cant}$ indicate the less importance of the multiband models for the anomalous TE effect in FeGe.

Additionally, the studies[54, 55] suggest that magnon drag thermopower substantially deceases at high temperatures owing to the dominant magnon-phonon scattering, and the unstable frustrated AFM magnons under high magnetic fields strongly suppresses the magnon-drag effect. As evidenced in Figure 1h and Figure S9c in SI, the thermopower behavior in FeGe deviates markedly from characteristic magnon-drag signatures, confirming its negligible role in FeGe's thermopower. Furthermore, both the observed magnetic field sensitivity and the breakdown of $T^3$ scaling in FeGe can collectively exclude any substantial phonon-drag contribution[56].[See Figure S9 and S10 in SI for more detailed analysis]. Other mechanism such as nontrivial spin texture may be involved[36, 39]. Whileas, in our current study in FeGe system with B ∥ ab-axis, no spin-flop transition is observed (See Figure S3a). Consequently, the observed large anomalous Nernst effect with the B ∥ ab-plane cannot be primarily attributed to such spin textures.

Secondly, different from the Hall effect in a metal with a finite value, the finite Nernst effect generally implies an unusual origin, like magnon, pseudo-gap, or Berry phase from time-reversal symmetry broken[17, 19, 23]. In addition, the anomalous Hall conductivity and thermoelectric conductivity can be expressed as $\sigma_{ij} = -\frac{e^2}{\hbar} \int \frac{d^3k}{(2\pi)^3} f_{nk}\Omega_B(k)$, and $\alpha_{ij} = -\frac{e^2}{T\hbar} \int \frac{d^3k}{(2\pi)^3} s_{nk}\Omega_B(k)$, where $f_{nk}$ is the Fermi-Dirac distribution function, $\Omega_B(k)$ represents the Berry curvature, and $s_{nk} = (\epsilon_{nk} - \mu)f_{nk} + k_B T ln[1 + e^{-\beta(\epsilon_{nk}-\mu)}]$. Therefore, although both AHE and ANE originate from the Berry curvature in momentum space, AHE is determined by the sum of the Berry curvature over all the occupied bands, and the ANE is denoted by the energy derivative of the Hall conductivity at the Fermi energy. Furthermore, the Nernst effect provides additional information about the entropy and the velocities of the electrons near the Fermi surface. Thus, the NE is more sensitive and effective in reflecting the intrinsic mechanism of Fermi surface changes, topological spin textures and charge orders.

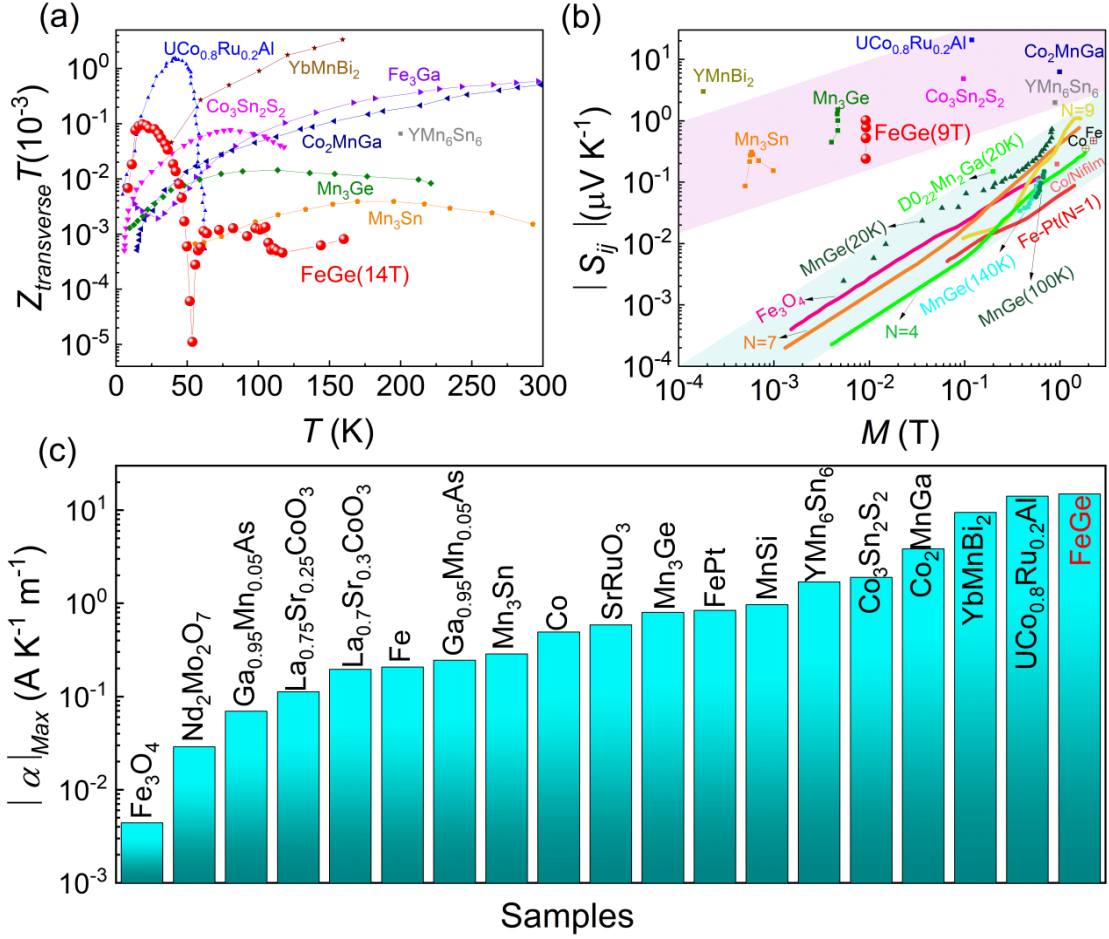

Figure 4 Comparison of Nernst- power factor (a) for FeGe and other topological magnets and antiferromagnets with large ANE thermopowers. (b) Magnetization scaling plot for the maximum ANE in conventional and topological FM/AFM materials. The recent ANE of topological magnets in the pink region is beyond the values reported for conventional magnetic materials (blue region), whose ANE values are proportional to magnetization. (c) Comparison of the maximum value $\alpha^A_{ij}$ of FeGe as well as those of reported magnetic materials with large ANE.

The transverse Nernst thermoelectric figure of merit $ZT = S_{zx}^2 \sigma T / k_{zz}$ in FeGe also shows a comparable value to other topological magnets, as shown in Fig. 4a, although it is still much lower than that of traditional longitudinal thermopower. However, this finding still is a promising progress towards the use of the transverse TE applications in Kagome AFM materials which possess highly dispersive Dirac bands together with magnetic bands to significantly enhance the TE performance. To further understand the mechanism of ANE in FeGe and compare its difference with other topological AFM and FM semimetals and conventional FM materials, the anomalous Nernst signal $S_{ij}$ as a function of magnetization is plotted in Figure 4b. In conventional ferromagnets, the anomalous Nernst signal $S^A_{ij} \propto N^A_{ij} \mu_0 M$ scales a linear relationship with magnetization, where $N^A_{ij}$ represents the anomalous Nernst coefficient that has rather small values from 0.05 to 1 $\mu V/K\,T$. Similar to most of topological FM and AFM, the $S_{zx}$ of FeGe is also far beyond the linear scaling relationship, which is almost one or two orders of magnitude smaller than those of conventional ferromagnets. This indicates that the origin of ANE in FeGe must be different from the conventional FM, but partially may be in common with topological magnets. Here, the large ANE can be attributed to a nonzero Berry curvature near the Fermi level because of the topological spin structures[46, 57]. Compared with other ferromagnets and topological magnets, the anomalous thermoelectric conductivity $\alpha^A_{zx}$ of 15 AK$^{-1}$m$^{-1}$ in FeGe (in Figure 4c) is found to be comparable to FM metal UCo$_{0.8}$Ru$_{0.2}$Al[23] and topological AFM material YbMnBi$_2$[35], but one or two orders of magnitude higher than those of the antiferromagnets Mn$_3$X (X = Sn and Ge)[20] and conventional FM.

## 3. Conclusion

In summary, we found that the Kagome AFM FeGe simultaneously exhibits some unconventional thermoelectric properties in the CAFM state, such as the large change in thermopower, and the anomalous Nernst effect as well as the largest transverse thermoelectric conductivity. Around $T_{cant}$, $S_{zz}$ is rather small, but field-induced change in thermopower $S_{zz}$ reaches almost $10^2$-$10^4$% at 14 T, which is comparable to well-known AFM thermoelectric materials. Meanwhile, a large ANE is observed with a maximum $S^A_{zx}$ of 1.2 μV/K at 20 K, which is higher than that of the Kagome canted-AFM Mn$_3$Sn/Ge. More importantly, a large anomalous transverse thermoelectric conductivity $\alpha^A_{zx}$ of 15 A K$^{-1}$m$^{-1}$ is obtained in AFM state. This value is the largest recorded in the well-known AFM materials, and is also comparable to the value reported for the ferromagnetic metal UCo$_{0.8}$Ru$_{0.2}$Al[23]. Considering the nontrivial spin structure at the entrance of CAFM state, the unconventional transverse thermoelectric properties can be attributed to a large nonzero intrinsic Berry curvature near the Fermi level. This work demonstrates the promise of thermoelectric materials in topological AFM without stray fields and also provides a novel way to enhance the thermoelectric effect by tuning the Fermi level.

## 4. Experimental methods

FeGe single crystals were grown by the chemical vapor transport (CVT) method, as reported in the previous literature[47]. The mixture of high-purity iron and germanium powders in the ratio of 1:1 with additional iodine as a transport agent was weighed and sealed in a quartz ampule under vacuum. The ampule was placed, heated up to 873 K, and kept at this temperature for two weeks in a two-zone furnace. Finally, the sample was quenched in the water from 873 K. The shiny as-grown FeGe crystals were further annealed at 643 K under an evacuated quartz ampule for 10 days. The obtained annealed crystals of FeGe are investigated in text.

The measuring crystals with both ab-plane and c-directional orientation are selected from the same batch. They are polished and then cut into bar shapes. The (magneto)resistivity and Hall resistivity measurements were simultaneously performed using the six-terminal method in a commercial 14T-cryogenic refrigerator. The magnetization was measured using a commercial PPMS-VSM. The thermos-conductivity, thermopower and Nernst effect were simultaneously performed with a one-heater–two-thermometer technique in a 14T-cryogenic refrigerator with a high-vacuum environment. The measuring configuration is shown in Fig. 1(b). The longitudinal measurements involve the resistivity and the thermopower, where the electric voltage is parallel to electrical current or heat flow. The corresponding physical quantities are marked with the subscript *zz*. And the transverse measurements include the Hall effect and the Nernst effect, where the electric voltage is normal to the electrical current or the thermal gradient. Correspondingly, the physical parameters are marked with the subscript zx.

For the thermoelectric effect measurements, two thermocouples were employed to measure the temperature gradient generated by a small heater chip, and thermal contacts were accomplished via gold wires. The distance of the thermometers ($L_z$) is ~1-2 mm in our measurements. With the increase in sample temperature, the temperature gradient increases correspondingly by increasing the current of the resistivity heater. In the Nernst effect measurements, a slight misalignment of the voltage contacts can give rise to an extra contribution associated with the longitudinal signals (i.e., thermopower $S_{zz}$). However, the Nernst effect is generally field-odd while the thermopower is field-even, therefore such a contribution can be canceled by reversing the direction of the magnetic field during measurements. The longitudinal resistivity(thermopower) and the transverse Hall (Nernst) were symmetrized (antisymmetrized) to exclude electrode misalignment. Here, the Nernst signal can be obtained through the formula: $S_{zx} = -E_x/\nabla_z T = -V_{xz} L_z/(L_x \delta T)$, where $L_x$ and $L_z$ are the distances of transverse voltage contacts and longitudinal temperature difference $\delta T$, and $V_{zx}$ is the transverse voltage difference.

## 5. Ab initio calculations.

The first-principles calculations were performed by using the code of the Vienna ab initio simulation package (VASP)[58, 59]. Following previous reports[39, 41], the exchange–correlation function was described using the generalized gradient approximation, following the Perdew-Burke-Ernzerhof parametrization scheme[60]. The cut-off energy was set at 550 eV to expand the wavefunctions into a plane-wave basis. The convergence criteria for the total energy and ionic forces were set to $1\times10^{-7}$ eV and –0.01 eV Å$^{-1}$, respectively. The A–type antiferromagnetic order along the $c$ axis at the Fe sites, a canted spin structure and spin-orbit coupling (SOC) were considered. To calculate the ANC, the ab initio density functional theory Bloch wavefunction was projected onto highly symmetric atomic-orbital-like Wannier functions[61] with a diagonal position operator using VASP code. To obtain precise Wannier functions, we included the outermost $s$ and $p$ orbitals for Ge and $d$ orbital for Fe to cover the full band overlap from the ab initio and Wannier functions. Based on the effective tight-binding model Hamiltonian, we performed the calculations for the ANC by the linear response Kubo formula approach in the clear limit as given by formula (1):

$$\alpha_{ij} = -\frac{e}{T\hbar}\sum_n \int \frac{d^3k}{(2\pi)^3} s_{nk}\Omega_B(k) \tag{1}$$

$$\Omega_B(k) = \Omega_{ij,n} = Im \sum_{n\neq m} \frac{\langle n|\frac{\partial H}{\partial k_i}|m\rangle\langle m|\frac{\partial H}{\partial k_j}|n\rangle - (i\leftrightarrow j)}{(\epsilon_{nk}-\epsilon_{mk})^2} \tag{2}$$

$$s_{nk} = (\epsilon_{nk} - \mu)f_{nk} + k_B T ln[1 + e^{-\beta(\epsilon_{nk}-\mu)}] \tag{3}$$

In Formula (1) and Formula (2), $\Omega_B(k) = \Omega_{ij,n}$ denotes the $ij$ component of the BC of the $n$th band, $n\rangle$ and $m\rangle$ are the eigenstates of $H$, and $\epsilon_{nk}$ and $\epsilon_{mk}$ are the corresponding eigenvalues. The $T$ in formula (3) is the actual temperature, in this study, the actual temperature was set to 20 K for all spin canting structures, $f_{nk}$ is the Fermi distribution, $\beta = 1/k_B T$ and $\mu$ is the Fermi level. To realize integrations over the Brillouin zone, a $k$ mesh of 240 × 240 × 240 points was used.

**Acknowledgments**

This work was supported by the Hangzhou Joint Fund of the Zhejiang Provincial Natural Science Foundation of China (under Grants No. LHZSZ24A040001) and the National Natural Science Foundation (NSF) of China (under Grants No. U1932155, 12204298, 12274019,12174334), H. Guo was funded by the Guangdong Basic and Applied Basic Research Foundation (Grant No. 2022B1515120020).

Corresponding authors: jinke_bao@hznu.edu.cn; sunyan@imr.ac.cn; zhuan@zju.edu.cn; yklee@hznu.edu.cn

**Author contributions**

Y.K.L. and Y.S. conceived the project. C. F. S. and J. K. B. grew the single crystals of FeGe. J.J. M., Y.Z.L., J.X.L., Y.W.Z., G.X.W., and Y.F.H. performed the transport properties measurement. Y.T.C. and H.J.G. performed the magnetization measurements. C.R., and Y.S. performed DFT calculations for FeGe. J.L.W., C.G. F., and J.H. D. discussed the data. Y.S. Z.A.X., and Y.K.L. wrote the paper with inputs from all co-authors.

**Competing interests**

The authors declare no competing interests.

Correspondence and requests for materials should be addressed to J.K. Bao, Y. Sun, Z.A. Xu, or Y.K. Li.